\documentstyle[aps,preprint]{revtex}
\begin{document}
\draft
\title{Dynamic versus thermodynamic approach to non-canonical equilibrium}
\author{Mauro Bologna$^{1}$, Michele Campisi$^{2}$, Paolo Grigolini$^{1,2,3}$}
\address{$^{1}$Center for Nonlinear Science, University of North Texas,\\
P.O. Box 305370, Denton, Texas 76203-5370 }
\address{$^{2}$Dipartimento di Fisica dell'Universit\`{a} di Pisa and\\
INFM \\
Piazza\\
Torricelli 2, 56127 Pisa, Italy }
\address{$^{3}$Istituto di Biofisica del Consiglio Nazionale delle\\
Ricerche, Area della Ricerca di Pisa, Via Alfieri 1, San Cataldo,\\
56010,\\
Ghezzano-Pisa, Italy }
\date{\today}
\maketitle

\begin{abstract}

We study the dynamic and thermodynamic origin of non-canonical equilibria,
and we discuss their connection with the generalized central
limit theorem and the micro-canonical Boltzmann principle.
We reach the conclusion that the zeroth law of thermodynamics
and the Boltzmann principle
are fulfilled thanks to an apparent fault turned into a
benefit: the dynamic approach
can only produce a truncated form of inverse power law equilibrium.
\end{abstract}

\pacs{02.50.Ey,05.20-y.05.40.Fb,05.70-a}

\section{Introduction}

This paper aims at establishing the most convenient theoretical framework
accounting for the emergence of non-canonical form of equilibrium.
As pointed out by the authors of Ref.\cite{juri}, one of the big
merits of the non-extensive thermodynamics of Tsallis\cite{tsallis}
is that of making popular the discussion of non-canonical equilibria
that a few years earlier might have caused the severe criticism  of the
referees\cite{juri}. We address the same problem from within the
dynamic perspective of Ref.\cite{bianucci} and we reach conclusions
that support theoretically the existence a form of non-canonical
equilibrium, in apparent accordance with the reasons of the
success of non-extensive thermodynamics\cite{success}. Actually, as we
shall see, our conclusions significantly depart from the tenets of
non-extensive thermodynamics \emph{\'{a} la } Tsallis, and
restate the importance of the ordinary Boltzmann
principle\cite{lebo1,lebo2,goldstein}, in accordance with the point of
view of Gross, illustrated in one of the papers of these
Proceedings\cite{gross}.

\section{Jaynes approaches to L\'{e}vy procesess}

It has been recently pointed out\cite{anna} that the adoption of the method
of entropy maximization {\em \'{a} la} Jaynes\cite{jaynes1,jaynes2}, with
the Shannon entropy replaced by the Tsallis entropy, does not yield directly
the L\'{e}vy distribution, but a probability density function $\Pi(x)$ that
for reiterated application of the convolution yields the stable L\'{e}vy
distribution. Here we show that, in principle, the method of entropy
maximization yields exactly the L\'{e}vy distribution, if a deliberate use
is made of the informationa nature of the Jaynes method\cite
{jaynes1,jaynes2}.

Let us discuss this way of proceeding in the Gaussian case, first. Let us
imagine that the problem to solve has to do with determining the most
probable form of a square summable function $f(x)$ belonging to the real
axis $[-\infty ,\infty ]$. We know that this function is symmetric around $%
x=0$ and we know the first two non-vanishing terms of its Taylor series
expansion about $x=0$,
\begin{equation}
f(x)=\frac{c_{0}}{2\pi }-\frac{c_{2}}{2\pi }x^{2}+\ldots .
\label{taylorseriesexpansion}
\end{equation}
It is convenient to stress that in this case the information available to us
is expressed by
\begin{equation}
f(0)=\frac{c_{0}}{2\pi }  \label{zeroth-orderderivative}
\end{equation}
and
\begin{equation}
\frac{d^{2}}{dx^{2}}f(x)|_{x=0}=-\frac{c_{2}}{2\pi }.
\label{second-orderderivative}
\end{equation}

It is apparently difficult to proceed with the method of entropy
maximization to guess the unknown form of this function. However, this is
made easy, if we move from the $x$-representation to the $k$- or
Fourier-representation of the function $f$. We notice, in other words, that
the information available to us can be expressed under the form of moment
constraints if we quit the representation of $f$ as a function of $x$, and
we focus on the Fourier transform of $f(x)$, denoted by $\hat{f}(k)$. In
fact, the knowledge of $c_{0}$ and $c_{2}$ can be used to set the
constraints
\begin{equation}  \label{zerothmoment}
\int dk \hat{f}(k) dk = c_{0}
\end{equation}
and
\begin{equation}  \label{secondmoment}
\int dk \hat{f}(k) k^{{2}}dk = c_{2}.
\end{equation}

According to the principle of entropy maximization\cite{mem}, we have
to look for the
maximum of the Shannon entropy
\begin{equation}
S[f] = - \int dk \hat{f}(k) ln \hat{f}(k),
\label{ordinaryentropyinthefourierspace}
\end{equation}
while taking into account the constraints of Eq. (\ref{zerothmoment})
and Eq.(\ref{secondmoment}) by means of the Lagrange multiplier method.
This yields
\begin{equation}
\hat{f}(k) = A e^{-\tau k^{2}},  \label{fouriertransformofagaussian}
\end{equation}
this resulting form being the Fourier transform of an ordinary Gaussian
distribution.

The derivation of the L\'{e}vy distribution from the proper extension of
these arguments is easy. We assume that the second piece of
information available to
us, rather than being expressed in terms of the second-order derivative of
Eq. (\ref{second-orderderivative}), is given by
\begin{equation}
\frac{d^{\alpha }}{d\left| x\right| ^{\alpha }}f(x)|_{x=0}=-\frac{c_{\alpha }%
}{2\pi },  \label{fractionalderivative}
\end{equation}
while the form of the first piece of information of Eq.
(\ref{zeroth-orderderivative}%
) is kept unchanged. Note that the symbol $\frac{d^{\alpha }}{d\left|
x\right| ^{\alpha }}$ denotes the symmetric fractional derivative\cite
{mainardi}, defined by its action on the Fourier space, which reads:
\begin{equation}
{\cal F}\left\{ \frac{d^{\alpha }}{d\left| x\right| ^{\alpha }}f\left(
x\right) \right\} =-\left| k\right| ^{\alpha }\widehat{f}\left( k\right) .
\label{explicitform}
\end{equation}
This means that in the Fourier space the constraint on the second moment of
Eq. ({\ref{secondmoment}) is now replaced by
\begin{equation}
\int dk\hat{f}(k)|k|^{\alpha }dk=c_{\alpha }.  \label{fractionalmoment}
\end{equation}
In this case the method of entropy maximization yields
\begin{equation}
\hat{f}(k)\propto e^{-\tau |k|^{\alpha }},  \label{hereweare}
\end{equation}
which is well known to be the Fourier transform of an $\alpha $-stable
L\'{e}vy process\cite{montroll}. }

In conclusion, we have solved the problem raised by Shlesinger
and Montroll\cite{montroll} about the derivation of L\'{e}vy processes from
a maximum entropy formalism, without departing from the extensive form of
Shannon entropy and without using strange logarithmic constraints, as
proposed by these authors\cite{montroll}. However, this interesting conclusion
does not establish the thermodynamic character  of L\'{e}vy
processes, since the information approach in this case might depart
from the Boltzmann principle\cite{lebo1,lebo2,goldstein,gross}.

\section{On the West-Seshadri statistics}
To set the dynamic approach in the proper perspective, we
remind the reader  about the deep connection between the Boltzmann
principle and dynamics established by the authors of
Ref.\cite{bianucci}. An oscillator, with coordinate $x$ and
velocity $v$ is coupled to a dynamic system, called {\em booster}
to emphasize the fact that no thermodynamic property is already used,
as it happens with the ordinary thermal
baths. The Hamiltonian coupling is given by $\kappa x \xi$, $\xi$ being the
coordinate of one of the particles of the booster, referred to as
{\em doorway} variable. The booster is a dynamic system in a condition of
strong chaos. We do not assign to it any thermodynamic property, but
a given energy E and we assume that the condition of strong chaos
makes it
reach the micro-canonical state. We write the Liouville equation of the
whole system, oscillator plus bath, and we derive from this equation, with a
projection method, the equation of motion of the oscillator. We prove
that, under the condition of time scale separation, this reduced
Liouville equation becomes equivalent to the Fokker-Planck equation,
thereby leading to a canonical form of equilibrium. No use of
thermodynamics has been done to reach this important conclusion, but
only dynamic properties have been invoked\cite{bianucci}. At this
stage we assume that the width
of the resulting oscillator equilibrium distribution can be identified with the
temperature $T$ of the booster, and we obtain\cite{bianucci}
\begin{equation}
     \label{bigbianucci}
     k_{B} T = [\frac{\partial}{\partial E} ln W(E)
     \frac{\partial}{\partial E} ln <\xi^{2}>
      + Re \hat \Phi_{\xi}(\omega)]^{-1},
       \end{equation}
       where $W(E)$ denotes the volume of booster multidimensional surface
       with energy $E$ and $\hat \Phi_{\xi}(\omega)$ is the
       Laplace transform of the correlation function of the doorway variable
       $\xi$, evaluated at $\omega$, the  oscillator frequency.
       We know that $W(E)$ is proportional to $E^{N}$,
       $N $ being the number of degrees of freedom of the booster. Thus, in
       the thermodynamic limit only the first term within the square
       bracket of Eq.(\ref{bigbianucci}) survives, thereby recovering
       the ordinary form of Boltzmann principle.
It is evident that this nice connection between dynamics and
thermodynamics is made possible by the fact that, using the language
of the advocates of non-extensive statistical
mechanics\cite{tsallis,success}, extensive conditions apply. This
is so because of the time scale separation between system of interest
and booster, resting on the fact that the function $\Phi_{\xi}(t)$ is
integrable. Furthermore, the oscillator of interest is coupled to only
one particle of the system. The coupling with all the particles of the
booster would prevent us from recovering the ordinary form of
Boltzmann principle. In this paper, we focus our attention on the
case where the former extensive condition is broken, and $\lim_{t
\rightarrow} \Phi_{\xi}(t) = const/t^{\beta}$, with $\beta < 1$.
We shall see that this dynamic condition
generates a special kind of non-canonical equilibrium. To
prepare the ground for this interesting form of non-canonical equilibrium,
illustrated in Section II B, we have to discuss first the case of free
diffusion.

\subsection{Free Diffusion}

We have now to address a crucial issue, that concerning the subtle
difference between L\'{e}vy flight and L\'{e}vy walk. Let us imagine a
random walker that walks according to the following prescription. At regular
intervals of time, $0,T,2T,\ldots $, we draw the random numbers $%
\eta _{1},\eta _{2},\eta _{3},\ldots $, playing the role of random
velocities. These random numbers are characterized by the probability
density $P(\eta )$ given by
\begin{equation}
P(\eta )=\frac{1}{2}(\mu -1)\frac{W^{\mu -1}}{(W+|\eta |)^{\mu }},
\label{densitydistributionofeta}
\end{equation}
the factor of $1/2$ taking into account the fact that the probability for
the walker to make jumps in the positive direction is the same as that of
making jumps in the negative direction. Note that our discussion rests on
the special case where the first moment of this distribution is finite,
while the second is infinite, namely,
\begin{equation}
2<\mu <3.  \label{levyinterval}
\end{equation}
The reason for this choice is transparent. The fact that $\mu <3$ ensures
that the second moment of the distribution is infinite, thereby making it
possible for us to depart from the Gaussian basin of attraction, which would
correspond to the case of canonical distribution. On the other hand, we have
to ensure that the first moment is finite, so as to establish, as we shall
see, a connection with a satisfactory dynamic approach.

Let us consider the case where the random drawing is carried out a given
number of times, $n$. This means that the random walker, moving along an
one-dimensional path, the $x$-axis, at ``time'' n is found in the position
\begin{equation}
x(n)=(\eta _{1}+\eta _{2}+\ldots \eta _{n}) T.  \label{positionattimen}
\end{equation}
Let us imagine now that this process is repeated an infinitely large number
of times, so as to build up a distribution density yielding the probability
for us to find the random walker at the position $x$ at time $n$, $p(x,n)$.
According to the generalized central limit theorem\cite{gnedenko}, the
Fourier transform, $\hat{p}(k,n)$, in the limiting condition of very large $%
n $'s, gets the analytical form
\begin{equation}
\hat{p}(k,n)=exp(-b|k|^{\alpha }n),  \label{gnedenkoprescription}
\end{equation}
where
\begin{equation}
\alpha \equiv \mu -1.
\end{equation}
In the physical condition corresponding to Eq.(\ref{levyinterval}), the
diffusion coefficient $b$ reads
\begin{equation}
b=-W^{\mu -1}sin\left( \frac{\pi \mu }{2}\right) \frac{\Gamma (3-\mu )}{(\mu
-2)}.  \label{diffusioncoefficient}
\end{equation}
It is important to point out that the asymptotic form of Eq. (\ref
{gnedenkoprescription}) is reached after applying the convolution procedure
for a finite number of times, $n_{crit}$, which is fixed to be of the order
of $10$\cite{gnedenko}. Thus, we make the assumption of considering a coarse
grained time scale $\bar n$, where also the infinitesimal change $d\bar n$
fits the important property $d\bar n >> n_{crit}$.

Let us now consider another random walk prescription. This has to do with
drawing the random numbers $\tau_{i}$'s, with the probability density $%
\psi(\tau)$ given by
\begin{equation}  \label{densitydistributionoftau}
\psi(\tau) = (\mu-1) \frac{T^{\mu-1}}{(T + \tau)^{\mu}}.
\end{equation}
This means that we can build up an infinite sequence of numbers, which are
then used to make a random walker walk with the following rules. At time $%
t=0 $, when the first random number, $\tau_{1}$, is selected, we also toss a
coin to decide whether the random walker has to move in the positive or in
the negative direction. The random walker walks with a velocity of constant
intensity $W$. Thus, tossing a coin serves the purpose of establishing
whether the velocity of the random walker is $W$ (head) or $-W$ (tail). This
condition of uniform motion lasts for a time interval of duration $\tau_{1}$%
. At the end of this condition of uniform motion, a new number, $\tau_{2}$,
is randomly drawn, and a new velocity direction is established by another
coin tossing. It is important to stress that the physical condition of Eq.(%
\ref{levyinterval}) corresponds to the non-vanishing mean time $<\tau>$,
whose explicit expression is:
\begin{equation}
<\tau> = \frac{T}{\mu - 2}.  \label{meanwaitingtime}
\end{equation}
We denote as {\em event} the joint process of random drawing of a number and
of coin tossing. We consider a time scale characterized by the
property $t >> <\tau>$.
It is evident that the number of events that occurred prior to a given time
$t$ is given by
\begin{equation}
n = \frac{t}{<\tau>}.  \label{numberofevents}
\end{equation}
Note that we set also the condition that $n>n_{crit}$, thereby implying that
rather than the absolute time $t$ we are considering the coarse time $\bar{t}
$ defined by
\begin{equation}
\bar{t}=\bar{n}<\tau >.  \label{realcoarsegrainingtime}
\end{equation}

If we adopt this coarse-grained time scale the position occupied by
the the particle according to Eq.(\ref{positionattimen}) becomes
indistinguishable from the random walker position prescribed by
\begin{equation}
x(\bar{t})=\xi _{1}\tau _{1}+\xi _{2}\tau _{2}+\ldots \xi _{\bar{n}}
\tau _{\bar{n}},  \label{position2}
\end{equation}
where $\xi _{i}$ denotes a stochastic variable with only two passible
values, either $W$ or $-W$, a variable that keeps the same sign for the
whole time duration of a time interval between two nearest-neighbor events.
We can also connect $P({\eta})$ to $\psi(\tau)$ as follows\cite{anna}
\begin{equation}
P(\eta )=\frac{T}{2W}\psi (\frac{\left| \eta \right| T}{W}).
\label{connection2}
\end{equation}
As pointed out in Ref.\cite{anna}, this has the effect of making
L\'{e}vy diffusion compatible with a dynamic and Hamiltonian derivation.

\subsection{Fluctuation-dissipation without a finite time scale}

In 1982 West and Seshadri\cite{west} made an interesting proposal to derive
a form of non-canonical equilibrium. This is described by the Langevin
equation
\begin{equation}
\frac{dx}{dt}=-\gamma x(t)+\eta _{L}(t).  \label{originalwork}
\end{equation}
Note that West and Seshadri\cite{west} assumed the variable $\eta $ to be a
L\'{e}vy stochastic process. This means that the continuous time $t$ of
their treatment must be identified with the coarse-grained time of Section
IIA. This is an important aspect that has fundamental consequences on the
dynamic realization of L\'{e}vy processes, and, consequently, on our dynamic
approach to non-canonical equilibrium. The equilibrium distribution emerging
from Eq. (\ref{originalwork}) can be easily evaluated by noticing that the
probability distribution $p(x,t)$ is driven by the following equation of
motion
\begin{equation}
\frac{d}{dt}p(x,t)=[\gamma \frac{d}{dx}+\frac{d^{\alpha }}{d\left| x\right|
^{\alpha }}]p(x,t).  \label{reasonablequationofmotion}
\end{equation}
This is the density picture equivalent to
Eq. (\ref{originalwork}). Both equations afford an attractive picture
of dynamics driven by both dissipation and fluctuation.
In the ordinary case the fluctuation process is
described by a second-order differential operator. The anomalous case here
under discussion forces us to replace the second-order derivative with a
fractional differential operator. The Fourier transform of Eq. (\ref
{reasonablequationofmotion}) obeys the time evolution equation
\begin{equation}
\frac{\partial }{\partial t}\hat{p}(k,t)=-b|k|^{\alpha }\hat{p}(k,t)-\gamma k%
\frac{\partial }{\partial k}\hat{p}(k,t).
\label{fouriertransformofthemotionequation}
\end{equation}
This equation yields the equilibrium distribution
\begin{equation}
\hat{p}(k,\infty )=exp\left( -\frac{b}{\alpha \gamma }|k|^{\alpha }\right) ,
\label{westseshadri}
\end{equation}
which we refer to as West-Seshadri (WS) non-canonical equilibrium. It is
important to point out that this form of equilibrium distribution has slow
tails inversely proportional to $|x|^{\mu }$. In other words, in the case
$2<\mu <3$ the equilibrium distribution keeps unchanged the power law nature
of the original fluctuation.

At this stage we have to fit the request of making our treatment
compatible with a Hamiltonian derivation\cite{bianucci}. As pointed
out in Ref.\cite{mario}, in accordance with the Hamiltonian
formulation advocated by Zaslavsky\cite{zaslavsky}, this condition
is fulfilled by replacing Eq.(\ref{originalwork}) with
\begin{equation}
\frac{dx} {dt} = - \gamma x(t) + \xi(t),  \label{dynamicwork}
\end{equation}
where $\xi(t)$ is the stochastic process described in Section IIA.
This has apparently the effect of making the WS statistics compatible
with a Hamiltonian derivation. However, Eq.(\ref{dynamicwork}) yields
an equilibrium that is not exactly equivalent to the predictions of
the WS statistics. In fact, it is straightforward to prove that the
trajectories moving always in the same direction, namely, those
exploring the largest distances from the origin, cannot overcome, in
the positive direction, the distance $x = x_{max} = \frac{W}{\gamma}$
and, in the negative direction, the distance $x = x_{min} = -
\frac{W}{\gamma}$.
In other words, in the long-time limit, $\frac{1}{\gamma} >> T$, we
realize a truncated L\'{e}vy equilibrium. In
Section V we shall see that this property becomes the key
ingredient to ensure thermalization between a Gauss and a L\'{e}vy
system.

\section{The micro-canonical Boltzmann principle and the generalized central
limit theorem}

The central idea of this section is borrowed from Rajagopal and Abe\cite
{rajagopalabekintchin}. In fact, these authors made the remarkable
observation that according to Khinchin\cite{khinchin} ordinary statistical
mechanics rests on the central limit theorem thereby making it
reasonable to expect that the non-extensive statistical mechanics is
based on
the Generalized Central Limit Theorem (GCLT)\cite{gnedenko}. We find this
observation very appealing and we want to discuss in this section its
consequences from within our perspective that, as mentioned in Section I, is
also based on the micro-canonical Boltzmann principle, in accordance with
other authors\cite{lebo1,lebo2,goldstein}.

Let us assume that we know the energy distribution of a small subsystem of a
macroscopic system that is assumed to obey the micro-canonical Boltzmann
principle. The small subsystem is denoted, for the sake of simplicity as
{\em particle}. Let us denote by $p(e)de$ the probability that the particle
energy $e$ is found in the small interval $[e,e+de]$.
We do not take position on the explicit form of this energy distribution. We
only make the assumption that
\begin{equation}
lim_{e \rightarrow \infty} p(e) = \frac{const}{e^{\nu + 1}},
\label{asynptoticenergy}
\end{equation}
with $0<\nu < 2$. This is compatible with both L\'{e}vy and Tsallis
statistics. Let us assume that the macroscopic system consists of $N$
independent particles and let us define the energy per particle, $\epsilon$:
\begin{equation}
\epsilon = \frac{\sum_{j =1}^{N} e_{j}}{N}  \label{energyperparticle}.
\end{equation}
Since the N particles are independent the ones from the others, we obtain
for $\epsilon$ the following expression
\begin{equation}
P_{N}(\epsilon) = \int \prod [de_{j}p(e_{j})] \delta \left(\epsilon - \frac{%
\sum_{j=1}^{N}}{N}\right).
\end{equation}
The characteristic function of the distribution $P_{N}(\epsilon))$, $\hat P%
_{N}(k)$, is related to the characteristic functions of the single particle
probability distribution $p(e_{j})$, $\hat p(e_{k})$, by
\begin{equation}
\hat P_{N}(k)) =\hat p^{N}(\frac{k}{N}).  \label{fromonetomany}
\end{equation}
The key aspect of the search we are doing is the following one. We study the
limiting case of $N \rightarrow \infty$ for the purpose of assessing if the
characteristic function becomes equivalent to the Fourier transform of a
delta of Dirac. If this happens, then the non-canonical equilibrium under
study is compatible with the micro-canonical principle. If it does not, the
non-canonical equilibrium distribution is found to be incompatible with the
micro-canonical principle.

We skip the discussion of the case $\nu < 1$, which is proved to be
incompatible with the micro-canonical condition,
and we focus our attention on the condition:

\begin{equation}
1<\nu <2.  \label{finitefirstmoment}
\end{equation}
In this case the first moment is finite and is denoted by the symbol $a$,
namely
\begin{equation}
a\equiv <e>=\int ep(e)de,  \label{energymeanvalue}
\end{equation}
and the L\'{e}vy-Gnedenko theorem\cite{gnedenko} affords an analytical
expression for the asymptotic distribution of the variable
\begin{equation}
u\equiv \frac{\sum_{k=1}^{N}e_{k}-Na}{N^{\frac{1}{\nu }}}.
\label{variablewithamean}
\end{equation}
We denote by $\hat L_{N}(k)$ the characteristic function of the
variable $u$, and we obtain
\begin{equation}
lim_{N\rightarrow \infty }\hat{L}_{N}(k)=exp\left[ ik\gamma -b|k|^{\nu
}\left( 1+\beta \frac{k}{|k|}tan(\frac{\pi \nu }{2})\right) \right]
\label{usualcharacteristicfunction2}
\end{equation}
and
\begin{equation}
\hat{L}_{N}(k))=e^{-ikaN^{1-\frac{1}{\nu }}}\hat{p}^{N}(\frac{k}{N^{\frac{1}{%
\nu }}}),  \label{fromonetomanywithscaling2}
\end{equation}
thereby yielding
\[
lim_{N\rightarrow \infty }\hat{p}^{N}(\frac{k}{N^{\frac{1}{\nu }}})=exp\left[
ik\gamma +ikaN^{1-\frac{1}{\nu }}-b|k|^{\nu }\left( 1+i\beta \frac{k}{\left|
k\right| }\tan \frac{\pi \nu }{2}\right) \right] ,
\]
and consequently:
\begin{equation}
lim_{N\rightarrow \infty }\hat{p}^{N}(\frac{k}{N})=exp\left[ ik\gamma N^{%
\frac{1}{\nu }-1}+ika-b|k|^{\nu }N^{1-\nu }\left( 1+i\beta \frac{k}{\left|
k\right| }\tan \frac{\pi \nu }{2}\right) \right]   \label{centralresult2}.
\end{equation}
In this case we see that increasing $N$ rather than a broader and broader
distribution makes the righ hand side of Eq.(\ref{centralresult2}) identical
to $exp(ika)$, namely the Fourier transform of a delta Dirac. Thus, this
case is compatible with the micro-canonical equilibrium.

At this stage it would be straigthforward to prove that the Tsallis
non-canonical equilibrium is compatible with the Boltzmann
principle provided that the entropic index $q$ fulfills the condition
$1 < q<7/5$. The WS statistics, on the contrary, would be incompatible
with the Boltzmann principle. However, as we shall see in Section V,
the dynamic approach to WS statistics, thanks to the fact
that the long tails are truncated, fits both the Boltzmann
principle and the zeroth law of thermodynamics at the same time.

\section{Thermal equilibrium between a non-canonical and a canonical system
and conclusions}

All the moments of a truncated L\'{e}vy process are finite, thereby
fitting the Khinchin prescriptions for an equilibrium distribution
to be compatible with the Boltzmann principle.
The same property makes this form of non-canonical equilibrium
compatible
with the zeroth principle of thermodynamics. In Ref.\cite
{juri} it has been shown that the finite second moment of the non-canonical
distribution can be evaluated analytically and its explicit expression is
\begin{equation}
<x^{2}(\infty)> = <\xi^{2}>T^{\mu -2} exp[\gamma (\mu-2) T] \frac{\gamma(3
-\mu, \gamma T)}{\Gamma^{3 -\mu}},
\end{equation}
where $\Gamma(\alpha,z)$ is the incomplete Gamma function.
This conclusion can be turned into a benefit. We think in fact that it
makes it possible to establish in a natural way the thermal
equilibrium between a system with L\'{e}vy statistics and one with
ordinary Gauss statistics. This naturally emerges from the theoretical
approach established years ago in Ref.\cite{faetti} to study the
process of heat tranfer from a hotter to a warmer system. Let us
imagine that a Gaussian oscillator, with temperature $T$, is weakly coupled
to a truncated L\'{e}vy process, whose temperature has to be assessed
using the Gauss system as a thermometer. The thermometer equilibrium is
shown\cite{faetti} to depend only on the second moment of the
L\'{e}vy system, thereby
making it possible to establish  naturally the thermalization between
the two systems, and also to measure the L\'{e}vy temperature by means
of the Gauss thermometer.
In conclusion, the adoption of the dynamic perspective of
Ref.\cite{bianucci}, extended to the case of a booster producing
fluctuations with infinite correlation time, yields a non-canonical
form of equilibrium, which is compatible with the Boltzmann principle
and with the zeroth principle of thermodynamics.


\begin{references}

\bibitem{juri}  M. Annunziato, P. Grigolini, and J. Riccardi, Phys. Rev. E
{\bf 61}, 4801 (2000).

\bibitem{tsallis}  C. Tsallis, J. Stat. Phys. {\bf 52}. 479 (1988).

\bibitem{bianucci}  M. Bianucci, R. Mannella, B.J. West, P. Grigolini, Phys.
Rev. E {\bf 51}, 3002 (1995).

\bibitem{success}  Nonextensive statistical mechanics aims at offering a
theoretical framework for systems with long-range interaction, long-range
memory, or fractal structure. Within this new theoretical breakdown
canonical equilbrium is singular in a more general form of equilibrium,
called a generalized canonical equilibrium\cite{tsallis}. The number of
advocates of this theoretical perspective keeps increasing and a
comprehensive list of references is currently available at
http://tsallis.cat.cbpf.br/biblio.htm.


\bibitem{lebo1}  J.L. Lebowitz, Rev. Mod. Phys. {\bf 71} S346 (1999).

\bibitem{lebo2}  J.L. Lebowitz, Physica A {\bf 263}, 516 (1999).

\bibitem{goldstein}  S. Goldstein, arXiv:cond-mat/0105242, 11 May 2001.

\bibitem{gross} D.H.E. Gross, arXiv:cond-mat/0106496.

\bibitem{anna}  M. Buiatti, P. Grigolini, A. Montagnini, Phys. Rev. Lett.
{\bf 82}, 3383 (1999).

\bibitem{jaynes1}  E.T. Jaynes, Phys. Rev {\bf 106}, 620 (1957).

\bibitem{jaynes2}  E.T. Jaynes, Phys. Rev. {\bf 108}, 171 (1957).

\bibitem{mem} N. Wu, {\em The Maximum Entropy Method}, Springer
Series In Information Science, Springer, Berlin (1997).

\bibitem{mainardi}  A.Chechkin, V. Gonchar, R.Gorenflo, F.Mainardi,
L.Tanatorov, arXiv:cond-mat/0012155 (2000).

\bibitem{montroll}  E.W. Montroll, F. Shlesinger, in {\em Nonequilibrium
Phenomena II: from Stochastics to Hydrodynamics}, edited by J.L. Lebowitz
and E.W. Montroll, North-Holland, Amsterdam (1984).

\bibitem{gnedenko}  B. V. Gnedenko, A.N. Kolmogorov, {\em Limit
Distributions for Sum of Independent Random Variables}, Addison Wesley,
Reading (1954).

\bibitem{west}  B.J. West and V. Seshadri, Physica A {\bf 196}, 203 (1982).


\bibitem{mario}  M. Annunziato, P. Grigolini, B.J. West,
Phys. Rev. E {\bf 64}, 011107 (2001)

\bibitem{zaslavsky} G. W.  Zaslavsky {\em Physics of Chaos in
Hamiltonian Systems},
Imperial College Press, London (1998).


\bibitem{rajagopalabekintchin}  S. Abe and A.K. Rajagopal, Europhys. Lett.
{\bf 52}, 610 (2000).

\bibitem{khinchin}  A.I. Khinchin, {\em Mathematical Foundations of
Statistical Mechanics} Dover Publications, Inc. New York (1949).


\bibitem{faetti}  S. Faetti, L. Fronzoni, P. Grigolini,
Phys. Rev. A {\bf 32}, 1150 (1985).
\end{references}
\end{document}